\documentclass{article}

        \usepackage{PRIMEarxiv}
        \usepackage[utf8]{inputenc}
        \usepackage[T1]{fontenc}
        \usepackage{url}
        \usepackage{booktabs}
        \usepackage{amsmath}
        \usepackage{amsfonts}
        \usepackage{amssymb}
        \usepackage{nicefrac}
        \usepackage{microtype}
        \usepackage{fancyhdr}
        \usepackage{graphicx}
        \usepackage{float}
        \usepackage{makecell}
        \usepackage{array}
        \usepackage{tabularx}
        \usepackage[nocompress]{cite}
        \makeatletter
        \renewcommand{\citepunct}{,\penalty\@m}
        \makeatother
        \usepackage{hyperref}

        \title{DSA Nonce Vulnerabilities: An Interactive Analysis}

        \author{
          Rundong Wei \\
          Hainan University \\
          Haikou, China \\
          \texttt{weirundong2026@hainanu.edu.cn} \\
          \And
          Xiaomei Tian \\
          Hengyang Normal University \\
          Hengyang, China \\
          \texttt{tianxm@hynu.edu.cn} \\
          \And
          Xiaoqi Li \\
          Hainan University \\
          Haikou, China \\
          \texttt{csxqli@ieee.org}
        }

\begin{document}
        \maketitle

        \begin{abstract}
        Digital signatures are fundamental to identity authentication and data integrity in cybersecurity, and the NIST-standardized Digital Signature Algorithm (DSA) frequently appears in the cryptography track of CTF competitions. However, DSA relies on number theory, modular arithmetic, and large-integer computation, making both the algorithm and its associated attacks difficult for beginners to follow. Conventional tools often expose only inputs and outputs, leaving the intermediate computations of signing, verification, and key-recovery attacks opaque. This paper presents a DSA signature analysis and visualisation platform tailored to CTF competitions. The platform provides three main capabilities: basic signature generation and verification, reproduction of common CTF attack methods, and dynamic visualisation of attack workflows. It covers three representative nonce vulnerabilities: nonce reuse, linear nonce leakage, and HNP-based lattice attacks. Stepwise displays and highlighted intermediate values make the underlying computations directly inspectable. Experiments show that the platform correctly reproduces the standard DSA workflow and all three attack scenarios.
        \end{abstract}


\section{Introduction}

As Internet technologies continue to advance and digital transformation deepens, challenges in cyberspace security are becoming increasingly prominent. Against this background, CTF competitions have become an important means of cultivating cybersecurity talent. By presenting participants with diverse security problems, CTF evaluates both technical competence and teamwork across areas such as cryptography, Web security, reverse engineering, binary exploitation, and mobile device security. Recent cybersecurity research has also investigated exploit and vulnerability analysis in DeFi protocols, smart-contract bytecode, and blockchain platforms through cross-contract, semantic, and structural methods \cite{LiDeFiTail2025,BuSmartBugBert2025,WuSolana2025,LiCOBRA2026}. Cryptography is a foundational component of cybersecurity and plays a central role in CTF competitions. Among related techniques, the Digital Signature Algorithm (DSA), proposed by the National Institute of Standards and Technology (NIST) in 1991, is a widely used public-key digital signature scheme for identity authentication and data integrity verification. DSA-related CTF challenges assess whether participants understand the algorithm's internal mechanisms and can identify weaknesses in practical implementations.

However, DSA involves substantial number-theoretic knowledge, and both signature generation and verification require extensive modular arithmetic and large-integer computation. For beginners, these mechanisms are difficult enough on their own, while the attack techniques that frequently appear in CTF competitions make the topic even harder to understand. Common attack patterns include nonce reuse, linear nonce leakage, and lattice attacks based on the Hidden Number Problem (HNP) and Lenstra--Lenstra--Lovasz (LLL) lattice basis reduction. Understanding and applying these techniques requires a solid mathematical foundation and familiarity with the relevant attack methodology. Most available DSA analysis tools provide basic signature generation and verification without intuitive visualisation support. Consequently, users cannot directly observe the internal execution process of DSA or its associated attacks. These tools also provide limited support for nonce-related vulnerabilities, so competitors often need to write substantial amounts of code themselves. Therefore, developing a DSA signature analysis and visualisation platform specifically for CTF competitions is both necessary and timely. The platform has three practical objectives: to expose the intermediate mathematics of DSA, to reproduce three nonce-related attack workflows from user-supplied parameters, and to provide an instructional environment for cryptography and CTF training.

This paper makes the following contributions: 
\begin{enumerate}
	\renewcommand{\labelenumi}{(\arabic{enumi})}
	\item We analyse DSA key generation, signing, verification, and three nonce-related attack models used in CTF challenges. 
	\item We present a requirements-driven architecture that integrates standard DSA operations, attack analysis, visualisation, and auxiliary tools. 
	\item We implement the platform using Tkinter, PyCryptodome, and SageMath, and evaluate it through functional and performance tests.
\end{enumerate}

\section{Related Work}

Classical DSA research covers standard implementations, multi-party signature generation, proactive deployment, and broad assessments of performance and security \cite{Zulfah2021,MacKenzie2004,Draelos1998,Lyu2025}. Application-oriented implementations combine DSA with encryption, hashing, steganography, and digital-certificate authentication \cite{Aufa2018,Nazal2019,Taneja2015,Yulianto2022,Rachmawanto2022}. Other studies optimise verification or investigate hybrid, comparative, lightweight, and educational implementations of digital signatures \cite{Al-Sewadi2021,Sarkar2012,Fitriani2021,item25,Fitri2025,Kharisma2026,Sirpal2026}. Beyond conventional DSA, public-key and signature mechanisms have also been examined in blockchain, public Wi-Fi, and key-derivation settings \cite{Saymanov2025,Jeon2017,Perone2017,Bouftass2015}. Research on quantum-resistant signatures includes alternative constructions, quantum-signature models, migration analyses, and post-quantum blockchain security \cite{Anshel2017,Amiri2015,Chen2025,Al-Janabi2025,Fernandez-Carames2024}. For ML-DSA, recent work studies threshold signing, application deployment, public-key management, and authentication systems \cite{Bienstock2026,Celi2026,Desai2026,Liu2026,Mitra2025}. Hardware-oriented work develops unified architectures, NTT accelerators, area-efficient execution, and arithmetic optimisations \cite{Truong2025,Kundi2024,Pu2025,Kurdi2025,Chen2026}. Performance and implementation studies also consider GPU acceleration, masking countermeasures, and customised processor support \cite{Khodanpur2026,Raj2026,Ye2025}. Implementation-security studies further evaluate side-channel leakage, resource-constrained deployment, Rowhammer exploitation, software-induced fault attacks, and standards-compliant signature splitting \cite{Dobias2026,Chhetri2026,Boy2026,Korensky2026,Kim2023}. Shparlinski showed that partial knowledge of DSA nonces can compromise the private key, providing the theoretical basis for lattice-assisted key recovery \cite{Shparlinski2003}. Dedicated DSA attack tools can reproduce individual vulnerabilities, but they generally provide limited support for inspecting the intermediate computations used during nonce and private-key recovery. SageMath provides large-integer arithmetic, matrix operations, and lattice basis reduction, making it suitable for cryptographic analysis involving HNP and LLL \cite{Hegde2023,Somsuk2023,Al-Absi2020}. These capabilities are widely used in scripts for nonce-leakage attacks. However, command-line implementations typically expose the constructed lattice and reduced vectors only through raw output, which can make errors in parameter preparation difficult to identify. CryptoHack and related online platforms combine cryptographic challenges with instructional material covering algorithms such as DSA, RSA, and ECC. Existing cryptographic teaching platforms also use dynamic demonstrations to explain standard encryption and signature procedures. CTF training platforms provide practical challenge environments, but visualisation of DSA nonce attacks and their intermediate mathematical steps remains limited. The platform presented in this paper addresses this gap by integrating standard DSA operations, three representative nonce attacks, and stepwise visualisation in one environment.

\section{Background and Attack Models}

\subsection{DSA Signing and Verification}

DSA is primarily used for digital signatures. Its security relies on the computational hardness of the discrete logarithm problem in finite fields, and it therefore provides signing and verification rather than encryption or decryption \cite{Johnson2021,Yang2022,Kazmirchuk2020}. DSA uses a public-private key pair. The private key is kept secret by the signer and is used only for signature generation, whereas the public key is openly available and is used to verify the authenticity and validity of a signature. Both signature generation and signature verification in DSA involve extensive modular arithmetic and large-integer computation.

\paragraph{(1) Key generation.}
DSA key generation proceeds as follows:
\begin{enumerate}
\item Choose a large prime $p$ whose bit length $L$ satisfies $L=64k$ within the required range. NIST FIPS 186-4 also permits approved parameter lengths outside this basic expression.
\item Select a large prime $q$ such that $q$ divides $(p-1)$.
\item Choose an integer $h$ with $1<h<p-1$, and compute $g=h^{(p-1)/q}\bmod p$ subject to $g\neq 1$.
\item Select the private key $x$ uniformly from $1\leq x<q$.
\item Compute the public key $y=g^x\bmod p$.
\end{enumerate}
The resulting key pair comprises the public parameters $(p,q,g,y)$ and the private key $x$.

\paragraph{(2) Signature generation.}
For a message $m$, DSA generates a signature as follows:
\begin{enumerate}
\item Compute the message digest $h=H(m)$.
\item Select a nonce $k$ such that $1<k<q$ and $\gcd(k,q)=1$.
\item Compute $r=(g^k\bmod p)\bmod q$.
\item Compute $s=k^{-1}(h+xr)\bmod q$, where $k^{-1}$ is the multiplicative inverse of $k$ modulo $q$.
\item If $r=0$ or $s=0$, select a new nonce and repeat Steps 3 and 4.
\end{enumerate}
The signature on $m$ is the pair $(r,s)$.

\paragraph{(3) Signature verification.}
Given a message $m$, a signature $(r,s)$, and the public key $(p,q,g,y)$, verification proceeds as follows:
\begin{enumerate}
\item Check that $0<r<q$ and $0<s<q$. Otherwise, reject the signature.
\item Compute the message digest $h=H(m)$.
\item Compute $w=s^{-1}\bmod q$.
\item Compute $u_1=hw\bmod q$ and $u_2=rw\bmod q$.
\item Compute $v=((g^{u_1}y^{u_2})\bmod p)\bmod q$.
\item Accept the signature if and only if $v=r$.
\end{enumerate}

\subsection{Nonce Reuse Attack}

If a signer accidentally reuses the same nonce \textit{k} when signing two different messages, an attacker can use the two signatures to recover the private key \textit{x}. Assume there are two messages with hash values $h_1$ and $h_2$ and corresponding signatures $(r_1, s_1)$ and $(r_2, s_2)$. If the same nonce $k$ is used for both signatures, then $r_1 = r_2 = r$.

The signing equations for the two signatures are

\begin{equation*}
s_1 = k^{-1}(h_1 + xr) \bmod q \tag{1}
\end{equation*}
\begin{equation*}
s_2 = k^{-1}(h_2 + xr) \bmod q \tag{2}
\end{equation*}

Subtracting Equation (2) from Equation (1) yields:

\begin{equation*}
s_1 - s_2 = k^{-1}(h_1 - h_2) \bmod q \tag{3}
\end{equation*}

Multiplying both sides of Equation (3) by k yields:

\begin{equation*}
k(s_1 - s_2) = h_1 - h_2 \bmod q \tag{4}
\end{equation*}

From Equation (4), the nonce \textit{k} can be solved as:

\begin{equation*}
k = (h_1 - h_2)(s_1 - s_2)^{-1} \bmod q \tag{5}
\end{equation*}

Once k has been obtained, substituting it into Equation (1) yields the private key \textit{x}:

\begin{equation*}
x = (s_1 k - h_1)r^{-1} \bmod q \tag{6}
\end{equation*}

\subsection{Linear Nonce Attack}

If the nonce values satisfy the linear relation $k_i = a k_{i-1} + b \bmod q$, where $a$ and $b$ are known constants and $i$ denotes the signature index, then an attacker can recover the private key $x$ from two signatures. Consider two signatures whose nonces satisfy $k_2 = a k_1 + b \bmod q$. Assume they are generated for the same message with hash value $h$, and let the corresponding signatures be $(r_1, s_1)$ and $(r_2, s_2)$.

The corresponding signing equations are

\begin{equation*}
s_1 k_1 = h + x r_1 \bmod q \tag{7}
\end{equation*}
\begin{equation*}
s_2 k_2 = h + x r_2 \bmod q \tag{8}
\end{equation*}

Substituting the corresponding nonce expressions into Equations (7) and (8) yields:

\begin{equation*}
s_1 k_1 = h + x r_1 \bmod q \tag{9}
\end{equation*}
\begin{equation*}
s_2(a k_1 + b) = h + x r_2 \bmod q \tag{10}
\end{equation*}

Solving Equations (9) and (10) simultaneously gives the required intermediate quantity and then the private key \textit{x}.

\begin{equation*}
k_1 = [h(r_1-r_2) - b r_1 s_2](a r_1 s_2 - s_1 r_2)^{-1} \bmod q \tag{11}
\end{equation*}
\begin{equation*}
x = (s_1 k_1 - h)r_1^{-1} \bmod q \tag{12}
\end{equation*}

\subsection{HNP-Based Lattice Attack}

HNP-based lattice attacks represent one of the central difficulties in DSA challenges within CTF competitions. In recent years, many researchers have proposed and refined dedicated countermeasures and solving strategies, which have significantly accelerated the analysis of such problems \cite{Shparlinski2003}. The Hidden Number Problem (HNP) considers a prime $q$ and a collection of signature pairs of the form $(a_i, b_i)$ satisfying $b_i = (a_i x + e_i) \bmod q$, where $x$ is the hidden value to be recovered and $e_i$ is a relatively small error term. The central goal is to infer $x$ from these constrained data points together with their modular arithmetic relations. In DSA, if the high-order bits of the nonce \textit{k} are known, the recovery problem can be transformed into an HNP instance, and the private key \textit{x} can then be recovered by applying LLL lattice basis reduction.

Rearranging the DSA signing equation gives

\begin{equation*}
sk = h + xr \bmod q \tag{13}
\end{equation*}

Multiplying both sides of Equation (13) by the required factor yields:

\begin{equation*}
k = s^{-1}h + s^{-1}rx \bmod q \tag{14}
\end{equation*}

Define $A_i=s_i^{-1}r_i\bmod q$ and $B_i=s_i^{-1}h_i\bmod q$. Equation (14) can then be written as

\begin{equation*}
k_i=A_i x+B_i\bmod q \tag{15}
\end{equation*}

Here, $x$ is the hidden quantity. An HNP instance can be solved by constructing a lattice basis and applying LLL reduction to obtain a short vector from which $x$ can be recovered. The resulting lattice basis is

\begin{equation*}
G =
\begin{pmatrix}
q & 0 & \cdots & 0 & 0 & 0 \\
0 & q & \cdots & 0 & 0 & 0 \\
\vdots & \vdots & \ddots & \vdots & \vdots & \vdots \\
0 & 0 & \cdots & q & 0 & 0 \\
A_1 & A_2 & \cdots & A_n & 1 & 0 \\
B_1 & B_2 & \cdots & B_n & 0 & K
\end{pmatrix} \tag{16}
\end{equation*}

In this setting, let $n$ denote the number of signatures and treat $K = 2^m$ as a fixed constant. Applying LLL reduction to matrix \textit{G} yields a relatively short vector whose second-to-last element is exactly the private key \textit{x}.

\section{Design and Implementation}

\subsection{Design Goals and Architecture}

The platform is intended for CTF participants and cybersecurity learners. Its design goals are to expose the internal operations of DSA, demonstrate nonce-related vulnerabilities, and support repeatable analysis through an interactive interface. The platform adopts the three-layer architecture shown in Figure 1. We separate presentation, task processing, and data computation to simplify module maintenance and later extension. A requirement survey based on questionnaires and user interviews revealed the following priorities:
\begin{enumerate}
\renewcommand{\labelenumi}{(\arabic{enumi})}
\setlength{\itemsep}{0.35em}
\setlength{\topsep}{0.35em}
\setlength{\parsep}{0pt}
\setlength{\parskip}{0pt}
\item Support DSA key generation, signing, verification, and representative CTF attack scenarios.
\item Present signing, verification, and key-recovery computations step by step, with security-critical parameters highlighted.
\item Provide a clear GUI that minimises configuration overhead.
\item Ensure responsive large-integer and lattice computations.
\item Provide editing utilities, one-click data clearing, and unified flag generation.
\end{enumerate}

\begin{figure}[H]
\centering
\includegraphics[width=0.92\textwidth]{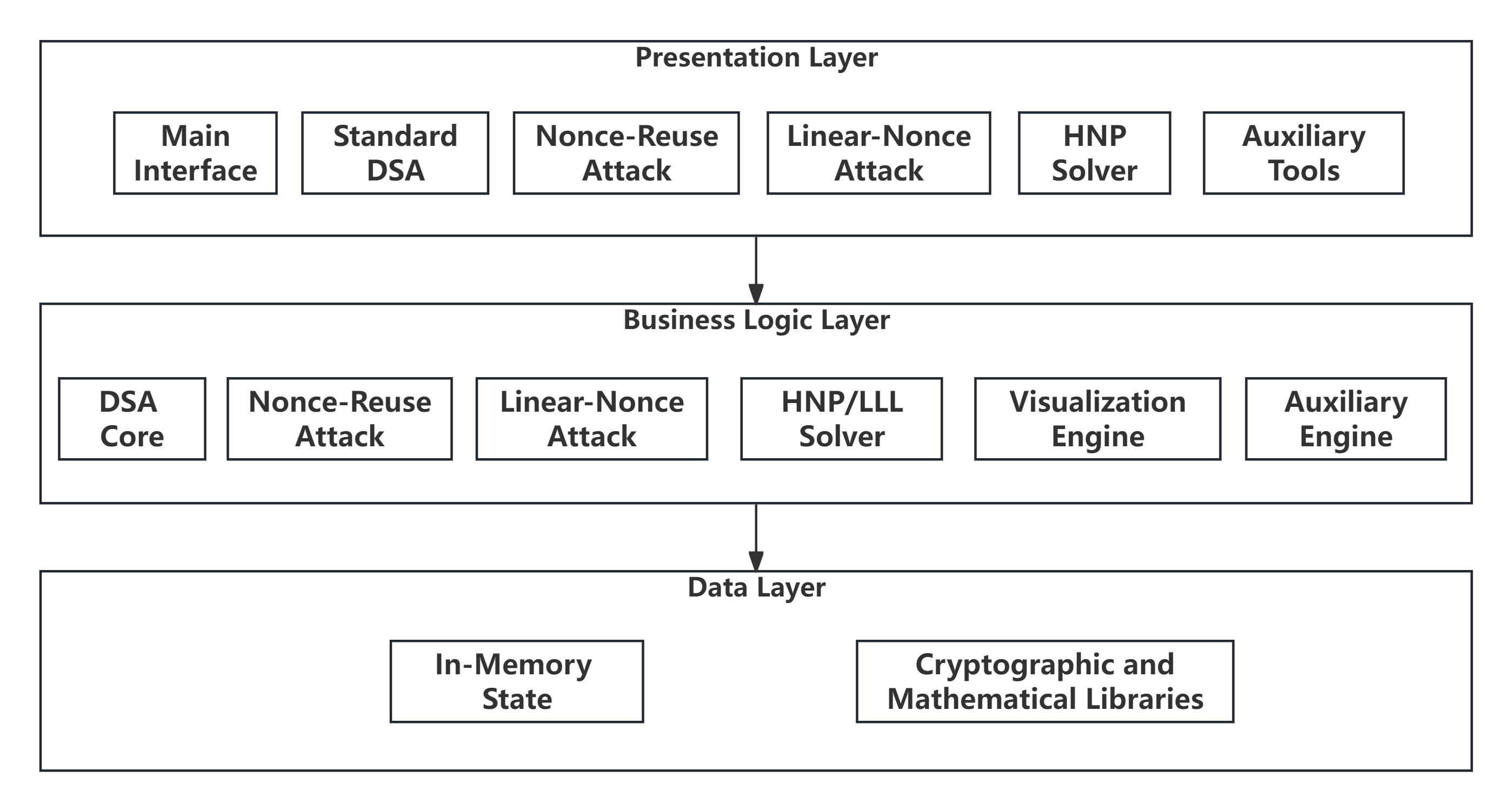}
\caption{Overall platform architecture.}
\end{figure}

Performance evaluation focuses on modular exponentiation and modular inversion for 1024-bit to 4096-bit integers, together with LLL reduction for HNP instances. The target latency is below 10 seconds for an HNP instance containing 10 signatures, while routine interface actions such as page transitions should complete within 0.5 seconds. Usability requirements emphasise a consistent layout, clear input guidance, and informative validation messages. The application should remain extensible to additional hash functions, nonce-attack models, and digital-signature algorithms. Python and Tkinter provide the application and GUI layers, PyCryptodome supplies the DSA primitives, and SageMath performs large-integer and lattice computations. This stack supports the required functionality without introducing a separate database or service layer. Tkinter is a built-in Python library for GUI development that supports basic interface components such as windows, buttons, labels, and input fields. Its simple and intuitive design makes it suitable for beginners, and its cross-platform support allows applications to run on Windows, Linux, and macOS. The platform is implemented in Python and uses the Tkinter GUI framework to build an interface that is both accessible and visually clear. PyCryptodome is dedicated to cryptographic applications and can be regarded as an improved successor to PyCrypto. It provides extensive implementations of symmetric and asymmetric encryption algorithms, hash functions, digital signature algorithms, and other core primitives. In this platform, PyCryptodome is used for DSA signature generation and related computations. Its DSA implementation follows NIST standards and provides a secure and stable basis for the required cryptographic operations. SageMath is an open-source mathematical computing platform that integrates tools such as NumPy, SciPy, SymPy, and GAP into a unified and efficient toolkit. It provides strong support for number-theoretic analysis, linear algebra, and more complex tasks such as lattice basis reduction. In this study, SageMath is used for large-integer computation, modular inverse calculation, and LLL-based lattice reduction, making it well suited to cryptographic and number-theoretic analysis \cite{Hegde2023,Somsuk2023,Al-Absi2020}. SageMath provides a stable implementation of the LLL algorithm that can solve lattice basis reduction problems efficiently and accurately, making it a suitable choice for this class of mathematical tasks.

\paragraph{Presentation layer.}
The Tkinter presentation layer handles user input, page navigation, and result display. It provides dedicated workspaces for standard DSA operations, nonce reuse, linear-nonce analysis, and HNP solving, together with a shared context menu for copy, cut, paste, and undo operations.

\paragraph{Business logic layer.}
The business logic layer dispatches requests to the DSA core, three attack modules, visualisation engine, and auxiliary functions. It coordinates key generation, signing, verification, nonce recovery, lattice construction, result formatting, and flag generation before returning the computed values to the active interface page.

\paragraph{Data layer.}
The data layer stores keys, signatures, and intermediate results in memory. PyCryptodome performs the cryptographic operations, while SageMath provides modular arithmetic and LLL lattice reduction. No persistent database is required for the current workflows.

\paragraph{Interface design.}

The interface follows four principles: visual consistency, low interaction overhead, clear separation between input and output, and support for later module expansion. A top toolbar provides direct access to standard DSA operations, nonce reuse analysis, linear-nonce analysis, HNP solving, and data-clearing functions. The content area below the toolbar is divided into parameter-input and result-display regions so that users can inspect the current operation without changing context. The DSA page contains regions for key generation, signature generation, signature display, verification, and results. The nonce reuse page separates the two signatures from the shared parameters and recovered values. The linear-nonce and HNP pages place parameters and controls on the left, signature sets on the right, and analysis results below. Scrollable regions accommodate larger signature collections without changing the overall layout.

\subsection{DSA Module}

The basic DSA module implements key generation, signature generation, and signature verification. It can either generate NIST-compliant parameters $(p,q,g)$ and the key pair $(x,y)$ or validate parameters supplied manually. For signing, the module hashes an input message with SHA-1 or SHA-256, supports normal, reused, and linearly related nonce modes, and displays both the nonce $k$ and the resulting signature $(r,s)$. For verification, it validates the signature range, reproduces the intermediate computations, and reports both successful and failed outcomes with the relevant reason. Internally, the key-generation submodule calls \texttt{DSA.generate()} for automatic generation and applies the expected mathematical checks to manually entered values. The signature-generation submodule hashes the message, derives $k$ according to the selected nonce mode, computes $(r,s)$, and sends each intermediate value to the visualisation module. The verification submodule hashes the message, checks the range of $(r,s)$, computes the verification value $v$, and accepts the signature when $v=r$.

The implementation follows the standard DSA workflow and uses PyCryptodome for NIST-compliant key generation, signing, and verification \cite{Hegde2023,Lyu2025}. It first checks whether the user supplied domain and key parameters. If the parameter fields are empty, \texttt{DSA.generate()} creates a 1024-bit parameter set $(p,q,g)$, after which the program obtains the private key $x$ and public key $y$. For signature generation, the program hashes the input message with SHA-1 or SHA-256 and derives the nonce according to the selected mode: random generation, reuse of a specified value, or a linear relation. It then computes $(r,s)$ and validates the result. For verification, the program reads the message and signature, computes the selected message digest, derives the intermediate verification values, and reports whether the signature is valid. Each stage forwards its formulas and numerical values to the visualisation module.

\subsection{Attack Module}

The nonce reuse attack module analyses two signatures produced with the same nonce $k$. It accepts $(p,q,g,y)$, two message-signature pairs, and a choice of SHA-1 or SHA-256. It checks whether the signatures satisfy the nonce-reuse condition, recovers $k$ and $x$, and presents the derivation step by step. The result view reports the recovered values, compares $x$ with a known key when available, and can convert the recovered key into a flag-compatible byte sequence. The linear-nonce attack module analyses signatures whose nonces satisfy a linear relation. It accepts $q$, the coefficients $a$ and $b$, and a user-managed set of message-signature entries. After hashing the messages with the selected algorithm, the module uses the first two signatures to derive the associated nonces and recover $x$. The interface exposes the main algebraic steps, displays the recovered values, and supports flag generation from the private key. The HNP module recovers the private key $x$ from multiple signatures by applying LLL lattice reduction. It accepts $q$ and a collection of message-signature entries, with SHA-1 and SHA-256 available for message hashing. It constructs the lattice basis, invokes SageMath's LLL implementation, extracts a candidate key, and verifies that candidate against every supplied signature. The matrix construction, reduction, and validation stages are displayed in sequence, and a validated key can be converted into the flag format. Together, the three modules connect each nonce vulnerability to the corresponding DSA signing equations and forward their intermediate values to the visualisation module.

\paragraph{Nonce reuse implementation.}

The nonce reuse implementation follows the derivation in Section 3.2. It reads $q$, two messages, and their signatures, computes both message digests, and then recovers $k$ and $x$ from the shared-$r$ equations.

\paragraph{Linear nonce implementation.}

The linear-nonce implementation reads $q$, the coefficients $a$ and $b$, and the message-signature set. It hashes the messages, selects two signatures, and evaluates the required numerators, denominators, and modular inverses to recover $k_1$, $k_2$, and finally $x$.

\paragraph{HNP implementation.}

The HNP implementation reads $q$, the selected hash algorithm, and multiple message-signature pairs. After computing the message digests, it derives the arrays $A$ and $B$, constructs the lattice basis, applies LLL reduction, and validates the recovered private key $x$.

\subsection{Visualisation and Auxiliary Functions}

The visualisation module presents each signing, verification, and attack computation in execution order, pairing the relevant formula with its numerical result. Colour-coded styles distinguish headings, intermediate steps, successful operations, warnings, formulas, and security-critical values such as $k$ and $x$. The result areas also support standard copy, cut, and paste operations. The auxiliary module provides a shared context menu, per-field undo history of up to 100 states, one-click clearing of the current page, and unified flag generation from recovered private keys. These functions are available throughout the Windows, Linux, and macOS interfaces rather than being tied to a single analysis page. The implementation organises each computation as an ordered stream of formulas and values written to predefined result regions. Tkinter text tags control colour, font family, and font size for each output category, while mouse-wheel scrolling keeps long traces accessible. In CTF workflows, the visual emphasis is placed on the nonce, recovered private key, attack condition, and key algebraic transformations rather than on decorative interface elements. The auxiliary module provides a range of supporting functions that make the platform more user-friendly and improve the overall user experience. Its structure is shown in Figure 2. These auxiliary functions are available across the functional pages rather than being isolated on a separate page.

\begin{figure}[htbp]
\centering
\includegraphics[width=0.9\textwidth]{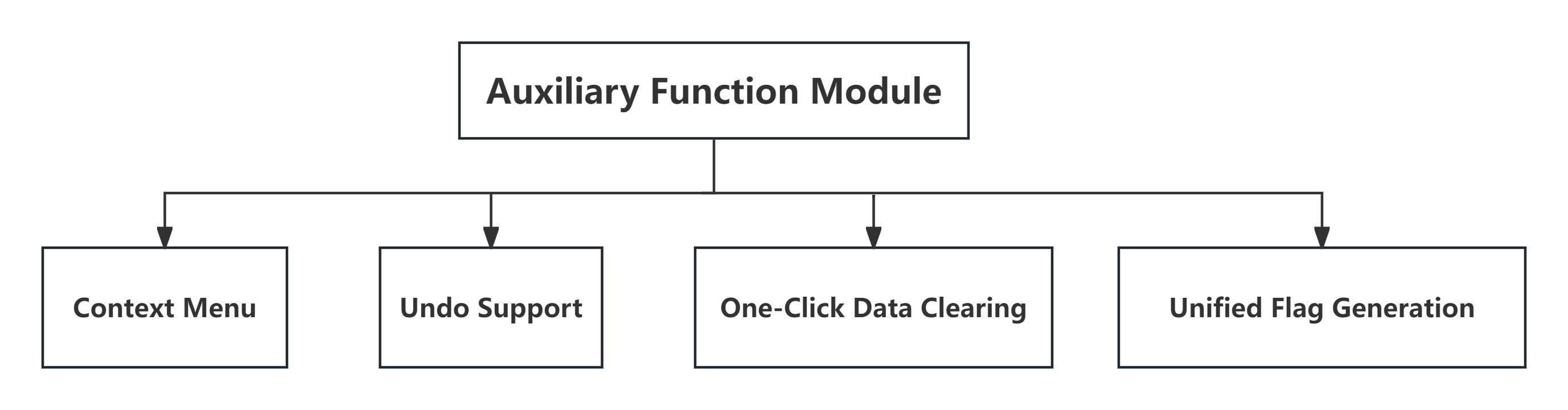}
\caption{Structure of the auxiliary function module.}
\end{figure}

The context-menu submodule binds copy, cut, paste, undo, and field-clearing commands to every input and text widget. Each input field maintains its own stack of previous states, allowing an undo command to restore the most recent value. One-click clearing removes the current page's inputs and outputs and resets the associated state variables. The flag-generation submodule stores the latest recovered private key, converts it to bytes, formats it according to the configured CTF convention, and writes the result to the active output panel. The \texttt{append\_result} method inserts content into the selected result panel, applies the requested style (for example, a heading, computation step, formula, success message, or error message), and moves the cursor to the latest output. All computation functions call \texttt{append\_result} to display formulas and numerical values in execution order. During signature generation, the trace contains the input message, message digest, nonce $k$, and signature values $(r,s)$. Security-critical values are assigned dedicated tags so that they remain visually distinct from explanatory text. The flag-generation function identifies the active result panel and converts the private key stored in \texttt{current\_x} into the required byte representation.

\section{Evaluation}

The evaluation verifies whether the platform satisfies its functional requirements and performance targets. All experiments were carried out in a unified hardware and software environment to ensure reproducibility, with detailed configurations provided in Table 1.

\begin{table}[H]
\centering
\caption{Experimental environment}
\small
\setlength{\tabcolsep}{3pt}
\renewcommand{\arraystretch}{1.08}
\begin{tabular}{@{}>{\raggedright\arraybackslash}p{0.24\textwidth}>{\raggedright\arraybackslash}p{0.70\textwidth}@{}}
\toprule
Item & Specification \\
\midrule
Processor & 11th Gen Intel(R) Core(TM) i5-1155G7 @ 2.50 GHz \\
Memory & 16 GB \\
Storage & 2 TB \\
Graphics & Intel(R) Iris(R) Xe Graphics 128 MB \\
Operating system & Windows 11 Home \\
Python & 3.13 \\
PyCryptodome & 25.0.1 \\
SageMath & 10.5 \\
\bottomrule
\end{tabular}
\end{table}

\paragraph{DSA functions.}

Basic DSA function testing is divided into three parts: key generation, signature generation, and signature verification. The corresponding results are shown in Table 2.

\begin{table}[H]
\centering
\caption{Basic DSA function test results}
\footnotesize
\setlength{\tabcolsep}{2.5pt}
\renewcommand{\arraystretch}{1.12}
\begin{tabular}{@{}>{\raggedright\arraybackslash}p{0.04\textwidth}>{\raggedright\arraybackslash}p{0.195\textwidth}>{\raggedright\arraybackslash}p{0.215\textwidth}>{\raggedright\arraybackslash}p{0.385\textwidth}>{\centering\arraybackslash}p{0.10\textwidth}@{}}
\toprule
No. & Test case & Test procedure & Expected result & Actual result \\
\midrule
1 & Automatic key generation & Click \texttt{Generate Keys} & Generate \textit{p}, \textit{q}, \textit{g}, \textit{x}, and \textit{y} & $\checkmark$ \\
2 & Input a valid key & Click \texttt{Generate Keys} & Validate the parameters & $\checkmark$ \\
3 & Input an invalid key & Click \texttt{Generate Keys} & Display an error prompt & $\checkmark$ \\
4 & Generate a signature normally & Click \texttt{Generate Signature} & Generate \textit{k}, \textit{r}, and \textit{s} & $\checkmark$ \\
5 & Generate a signature with reused \textit{k} & Select reused-\textit{k} mode & Generate a signature using the specified \textit{k} value & $\checkmark$ \\
6 & Generate a signature with linear \textit{k} & Select linear-\textit{k} mode & Generate a signature from linearly computed \textit{k} & $\checkmark$ \\
7 & Verify a valid signature & Verify with the same message & Display a successful verification process & $\checkmark$ \\
8 & Verify an invalid signature & Verify with a different message & Fail verification and report the reason & $\checkmark$ \\
\bottomrule
\end{tabular}
\end{table}

Visualisation testing examined the stepwise computation display and key-value highlighting. The platform reproduced the complete signing, verification, and attack workflows, paired each step with its formula and numerical result, highlighted the nonce $k$ and private key $x$ in red, and used distinct styles for different output categories. The resulting traces remained readable throughout the tested workflows. Attack-module testing focused on nonce reuse, linear-nonce attacks, and HNP solving. Standardised test datasets were used for evaluation, and the detailed results are listed in Table 3. The performance evaluation focused mainly on two aspects: large-integer computation speed and LLL lattice basis reduction, both of which are key bottlenecks in the platform's computational workflow. Each test case was executed ten times, and the results were averaged before reporting. During DSA signature generation, large-integer modular exponentiation is the main performance bottleneck. In the present evaluation, each test case was executed ten times in the specified hardware and software environment, and the arithmetic mean was reported in milliseconds (ms). The corresponding results are shown in Table 4. The LLL algorithm is central to solving HNP instances, and its computational speed directly affects the performance of the platform's attack module. In this experiment, the runtime of the platform on HNP instances with different numbers of signatures was measured. Each group was repeated 100 times, and the results were averaged, with runtimes reported in milliseconds.

\begin{table}[H]
\centering
\caption{Attack function test results}
\footnotesize
\setlength{\tabcolsep}{2.5pt}
\renewcommand{\arraystretch}{1.12}
\begin{tabular}{@{}>{\raggedright\arraybackslash}p{0.04\textwidth}>{\raggedright\arraybackslash}p{0.20\textwidth}>{\raggedright\arraybackslash}p{0.235\textwidth}>{\raggedright\arraybackslash}p{0.315\textwidth}>{\centering\arraybackslash}p{0.11\textwidth}@{}}
\toprule
No. & Test case & Input data & Expected result & Actual result \\
\midrule
1 & Successful reused-\textit{k} attack & Two signatures generated with the same \textit{k} value & Recover private key \textit{x} & $\checkmark$ \\
2 & Failed reused-\textit{k} attack & Two signatures with different \textit{r} values & Fail because $r_1 \ne r_2$ & $\checkmark$ \\
3 & Successful linear-\textit{k} attack & Input two signatures & Recover private key \textit{x} & $\checkmark$ \\
4 & Successful HNP solving & Input five signatures & Recover private key \textit{x} & $\checkmark$ \\
5 & Insufficient number of signatures & Input one signature & At least two signatures are required & $\checkmark$ \\
6 & Unified flag generation & Click the generate button & Output the flag & $\checkmark$ \\
\bottomrule
\end{tabular}
\end{table}

\begin{table}[H]
\centering
\caption{Large-integer computation performance results}
\small
\setlength{\tabcolsep}{3pt}
\renewcommand{\arraystretch}{1.08}
\begin{tabular}{@{}>{\raggedright\arraybackslash}p{0.24\textwidth}>{\raggedright\arraybackslash}p{0.36\textwidth}>{\raggedright\arraybackslash}p{0.34\textwidth}@{}}
\toprule
Big integer length (bits) & Average modular exponentiation time (ms) & Average modular inverse time (ms) \\
\midrule
1024 & 3.1 & 0.1 \\
\makecell[l]{2048 \\ 4096} & \makecell[l]{20.4 \\ 146.4} & \makecell[l]{0.4 \\ 1.1} \\
\bottomrule
\end{tabular}
\end{table}

\section{Discussion and Limitations}

The functional tests show that all evaluated modules satisfy the specified workflows. The visualisation traces expose the signing, verification, and key-recovery computations, while the auxiliary functions support routine editing and analysis tasks. Performance measurements indicate that the platform handles the 1024-bit and 2048-bit large-integer operations used in the test cases efficiently. For an HNP instance containing 10 signatures, the average solving time was approximately 2.0 ms. Testing also identified two interface-level issues: some error messages lacked detail, and layout consistency varied slightly across screen resolutions. The implementation was adjusted to improve error reporting, layout consistency, and interaction stability. The evaluation is limited to the attack models, parameter ranges, and test data implemented in the current platform. In particular, the HNP experiments use a specific leakage model and do not cover every form of biased or partially exposed nonce. Functional correctness was assessed through black-box test cases rather than a formal verification of the implementation. In addition, no controlled user study was conducted. Therefore, the platform's educational usefulness should be interpreted as a design objective rather than a quantitatively validated learning outcome. Penetration testing similarly treats offensive techniques as controlled evaluations aimed at improving system security \cite{ZhangPenTest2025}. The attack functions are intended for CTF competitions, authorised security testing, and cryptography education. They should not be applied to signatures or systems without explicit permission.

\section{Conclusion}

This paper presents an interactive platform for analysing and visualising DSA nonce vulnerabilities in CTF competitions. The Python implementation combines a Tkinter GUI, PyCryptodome, and SageMath to support key generation, signing, verification, nonce reuse attacks, linear-nonce attacks, and HNP-based key recovery with LLL reduction. Stepwise traces connect each recovered value to the corresponding DSA equation, while the shared editing and flag-generation functions support practical challenge workflows. Functional tests covered the basic DSA operations, all three attack modules, visualisation, and auxiliary functions. Performance tests measured large-integer arithmetic and LLL reduction without changing the attack models or parameter sets described in this paper. Within this scope, the platform achieves the expected results and provides readable intermediate computations.

\section*{Acknowledgments}
AI-based tools are used for language polishing during manuscript preparation.

\bibliographystyle{unsrt}
{\small
\bibliography{references}
}
        \end{document}